\documentclass[prb,twocolumn,amsmath,amssymb]{revtex4}

\usepackage{color}
\usepackage{bm}
\usepackage{dcolumn}
\usepackage[dvips]{graphicx}

\usepackage{ulem}

\begin{document}

\preprint{preprint}

\title{
Streda Formula for the Hofstadter--Wilson--Dirac Model
in Two and Four Dimensions
}

\author{Takahiro Fukui}
\author{Takanori Fujiwara}
\affiliation{Department of Physics, Ibaraki University, Mito 310-8512, Japan}

\date{\today}

\begin{abstract}
We rederive the spectral asymmetry of the Wilson--Dirac model in external fields,
paying attention to the Chern number due to the Berry connection.
We interpret the smooth part of the spectral asymmetry as the Streda formula that is originally   
derived for the two-dimensional quantum Hall effect (QHE).
We show by numerical calculations 
that the Streda formula reproduces the known first and second Chern numbers 
in a weak magnetic field limit. 
We conjecture that the Streda formula
is valid even for stronger fields and for more generic systems in higher dimensions. 
\end{abstract}

\pacs{
}

\maketitle

\section{Introduction}

Some symmetries in classical systems are broken in corresponding quantized systems.
\cite{Adler:1969aa,Bell:1969aa,Adler:1969ab,Bardeen:1969aa}
These phenomena are called anomalies in field theories, and have played an important 
role in particle physics. 
The chiral anomaly successfully explained the color degrees 
of freedom of quarks before the advent of QCD,
and the anomaly cancellation\cite{Bouchiat:1972aa,Gross:1972aa} provides 
a powerful guideline in unified gauge theories. It is well known that they 
are closely related with the topological structures of underlying gauge 
theories as summarized in the index theorem.
Topological aspects of anomalies imply 
their universality in physics. Indeed, 
the QHE can be interpreted \cite{Ishikawa:1985uq} as the parity anomaly of 
the Dirac fermion in three space-time dimensions.\cite{Redlich:1984kx}
According to the development of topological insulators, 
there is a growing interest in various kinds of anomalies in condensed matter physics.
\cite{Qi:2008aa,Ryu:2012aa}
Recently, topological phases in gapless systems, called Weyl and Dirac semimetals, have been proposed,
\cite{Murakami:2007aa,Burkov:2011fk,Young:2012aa}
and observed experimentally.\cite{Wang:2012aa,Xu:2015aa}
This motivates us to reconsider the experimental observations of the chiral anomaly of Weyl fermions
\cite{Son:2013aa,1605.08445,Ominato:2016aa,Reis:2016aa}
in a crystal proposed long ago.\cite{Nielsen:1983aa}

In this paper, we investigate the Wilson--Dirac model in strong external fields, 
which is referred to as 
Hofstadter--Wilson--Dirac (HWD) model, defined on the Euclidean
two- and four-dimensional spaces.
One of the purposes of this paper is to show an intimate relationship between the chiral anomaly of the Dirac fermion 
in particle physics and
the topological insulating phase in condensed matter physics. Indeed, Qi {\it et al.} have studied the same model
to construct the theory of the topological insulator.\cite{Qi:2008aa}
In Sect. \ref{s:CAL}, we briefly introduce the recent development of lattice gauge theory 
concerning the chiral anomaly on the lattice. 
Here, the spectral asymmetry plays a key role.
In Sect. \ref{s:CLSA}, we then rederive \cite{1126-6708-2002-09-025,Qi:2008aa} the smooth part of
the spectral asymmetry, paying attention to how the Chern number due to the Berry curvature appears.
We find that the relationship between the spectral asymmetry and the chiral anomaly is quite similar 
to the Streda formula, \cite{Streda:1982aa} 
which represents the Hall conductivity (and thus the first Chern number)
as the number of occupied states in the QHE.
Thus, the chiral anomaly represented by the spectral asymmetry that we discuss in this paper
is also referred to as the Streda formula.
Remarkably, the Streda formula enables us to compute the Chern number using eigenvalues only, 
without using eigenstates. 
It turns out that the Streda formula is valid even in four dimensions, and it serves as an efficient 
tool for computing the second Chern number related with the chiral anomaly.  
In Sects. \ref{s:TDHWDS} and \ref{s:FDHWDS}, 
we numerically show that the Streda formula reproduces the known Chern numbers of the HWD model
in the weak field limit
in not only two dimensions but also four dimensions. 
We finally conjecture in Sect. \ref{s:GSC}
that the Streda formula holds more generic systems even in four dimensions, and numerically 
suggest second Chern numbers
for several Landau levels formed in a simple tight-binding model. 

\section{Hofstadter--Wilson--Dirac model}
\label{s:HWD}
The Hamiltonian in $d$-dimensions is defined as
\begin{alignat}1
H=&\frac{-it}{2}\sum_{\mu=1}^d \sum_j\left(e^{-i\phi_{\mu,j}}c_j^\dagger\gamma_\mu c_{j+\hat\mu}-h.c.\right)
+m\sum_j c_j^\dagger\gamma_{d+1}c_j
\nonumber\\
&
+\frac{b}{2}\sum_{\mu=1}^d\sum_j 
\left(e^{-i\phi_{\mu,j}}c_j^\dagger\gamma_{d+1} c_{j+\hat\mu}
+h.c.-2c_j^\dagger\gamma_{d+1}c_j\right)
\nonumber\\
\equiv& \sum_{i,j}c^\dagger_i{\cal H}_{ij}c_j ,
\label{WDHam}
\end{alignat}
where $\hat\mu$ is the unit vector in the $\mu$-th ($\mu=1,\cdots,d$) 
direction and the $\gamma$-matrices
mean  for $d=2$ the Pauli matrices
$\gamma_1=\sigma_1,\gamma_2=\sigma_2,\gamma_3=\sigma_3$,
while they mean  for $d=4$ the standard hermitian $4\times4$ $\gamma$-matrices.
The phase $\phi_{\mu,j}$ stand for an external gauge field, which will be 
specified momentarily.
We investigate the spectral flow as a function of a uniform magnetic field, 
i.e., the Hofstadter butterfly.
Here, the spectral asymmetry plays a key role in extracting the topological nature from the butterfly, which is
defined as
\begin{alignat}1
\eta\equiv \frac{1}{2}{\rm Tr }\frac{{\cal H}}{\sqrt{{\cal H}^2}} =\frac{N_+-N_-}{2},
\label{Eta}
\end{alignat}
where ${\rm Tr}$ implies the trace over the $\gamma$-matrices as well as the space $j$, and 
$N_\pm$ stands for the numbers of the positive and negative energy states of ${\cal H}$.

\subsection{Chiral anomaly on the lattice}
\label{s:CAL}
The spectral asymmetry (\ref{Eta}) is closely related with chiral symmetry on the lattice:
The naive chiral symmetry is broken by the Wilson term
even when $m=0$ in Eq. (\ref{WDHam}). 
Without the Wilson term, the fermion  suffers from 
the doubling and the chiral anomaly cancels out among the fermion and the doublers. 
\cite{Nielsen:1981aa,Nielsen:1981ab}
It is known, however, that we can define the chiral invariant lattice fermion 
using the Dirac operator $D$ satisfying the Ginsparg-Wilson relation \cite{Ginsparg:1982aa}
$\gamma_{d+1}D+D\gamma_{d+1}=aD\gamma_{d+1}D$, where $a$ is the lattice constant. 
This leads to chiral symmetry on the lattice: $D\gamma_{d+1}(1-aD/2)+(1-aD/2)\gamma_{d+1}D=0$.\cite{Luscher:1998aa}
The chiral anomaly is then given by \cite{Luscher:1998aa,Kikukawa:1999aa,Suzuki01071999}
${\cal A}_x=\mathrm{tr}\gamma_{d+1}\left(1-\frac{a}{2}D\right)_{x,x}$, 
where $x=aj$ stands for the lattice coordinates and $\mathrm{tr}$ implies the trace over 
the $\gamma$-matrices. Such $D$ can be explicitly found as the overlap Dirac operator 
\cite{Neuberger:1998ab,Neuberger:1998aa}
$  D=\frac{1}{a}\left(1-\gamma_{d+1}\frac{\cal H}{\sqrt{{\cal H}^2}}\right)$.
The spectral asymmetry (\ref{Eta}) can be related with the chiral anomaly by
the lattice version of the index theorem 
$  \eta=\sum_x{\cal A}_x$.\cite{Adams:2002aa}
Thus, $\eta$ is topological.

\subsection{Continuum limit of the spectral asymmetry}
\label{s:CLSA}
If a state flows across zero energy, $\eta$ changes discontinuously.
Therefore, $\eta$ is composed of the smooth part $\bar\eta$ and the discontinuous part $\eta_{\rm d}$,
\begin{alignat}1
\eta=\bar\eta +\eta_{\rm d} .
\label{EtaTwo}
\end{alignat}
Below, we derive $\bar\eta$, according to Ref. \cite{1126-6708-2002-09-025},
especially paying attention to how the Chern number associated with the Berry curvature appears. 
To this end, we
calculate $\eta$ up to $a^d$.
Let $\nabla$ and $\nabla^*$ be the forward and backward difference operators, respectively, defined as
\begin{alignat*}1
a\nabla_{\mu}c_x&=e^{aA_\mu(x)}c_{x+a\hat\mu}-c_x,
\nonumber\\
a\nabla_{\mu}^*c_x&=c_x-e^{-aA_\mu(x-a\hat\mu)}c_{x-a\hat\mu} ,
\end{alignat*}
where $A_\mu(x)\equiv -i\phi_{\mu,j}/a$ is purely imaginary.
The fermion operators have been labeled by $x=aj$.  
Note that $\sum_{x} c_{x}^\dagger (a\nabla_\mu c_x)=\sum_{x,y}c_x^\dagger (a\nabla_{\mu}\delta_{x,y})c_y$,
where $a\nabla_\mu$ in the r.h.s operates on $x$ of $\delta_{x,y}$. 
Then, as the first quantized form, ${\cal H}$ in Eq. (\ref{WDHam}) is denoted as
\begin{alignat*}1
{\cal H}_{x,y}=\hat{\cal H}\delta_{x,y}\equiv\gamma_I\hat X_I\delta_{x,y},
\end{alignat*}
where $\hat X_I$ operates to $x$ and is defined as
\begin{alignat}1
\hat{X}_I=\left\{
\begin{array}{ll}\displaystyle
\frac{-it}{2}a(\nabla_\mu+\nabla_\mu^*) & (I=\mu)
\\\displaystyle
m+\frac{b}{2}\sum_{\mu=1}^d a(\nabla_\mu-\nabla_\mu^*) & (I={d+1})
\end{array}
\right. .
\label{XA}
\end{alignat}
Noting 
$\delta_{x,y}=\int_{-\pi}^\pi e^{i\frac{k}{a}\cdot(x-y)}\frac{d^dk}{(2\pi)^d}$,
we have the representation of $\eta$ suited for deriving the continuum limit,
\begin{alignat}1
\eta&=
\frac{1}{2}\sum_x {\rm tr}
\left(\frac{{\cal H}}{\sqrt{{\cal H}^2}}\right)_{x,x}
\nonumber\\&
=\frac{1}{2}\sum_x\int_{-\pi}^\pi e^{-i\frac{k}{a}\cdot x}{\rm tr}
\frac{\hat{\cal H}}{\sqrt{\hat{\cal H}^2}}e^{i\frac{k}{a}\cdot x}
\frac{d^dk}{(2\pi)^d}.
\label{Eta2}
\end{alignat}

Next, let us calculate the $\eta$ in the small $a$ limit.
Note that
\begin{alignat}1
e^{-i\frac{k}{a}\cdot x}a\nabla_{\mu}e^{i\frac{k}{a}\cdot x}
&=e^{ik_\mu}a\nabla_\mu+e^{ik_\mu}-1
\nonumber\\
&= iK_\mu+\delta_\nu K_\mu aD_\nu +O(a^2),
\label{DifDif1}
\end{alignat}
where $D_\mu$ is the covariant derivative
defined as
\begin{alignat*}1
D_\mu=\partial_\mu+A_\mu,
\end{alignat*}
with $\partial_\mu=\partial/\partial x_\mu$, and $K_\mu$ is an exponentiated momentum $k_\mu$ defined as 
$iK_\mu\equiv e^{ik_\mu}-1$. The derivative with respect to $k_\mu$ is denoted as $\delta_\mu$ such that
$\delta_\nu K_\mu\equiv\partial K_\mu/\partial k_\nu=\delta_{\mu\nu}e^{ik_\mu}$ to distinguish it from 
$\partial_\mu$.
Likewise, we have
\begin{alignat}1
e^{-i\frac{k}{a}\cdot x}a\nabla_{\mu}^*e^{i\frac{k}{a}\cdot x}
&=iK_\mu^*+\delta_\nu K_\mu^* aD_\nu +O(a^2) .
\label{DifDif2}
\end{alignat}
Using Eqs. (\ref{DifDif1}) and (\ref{DifDif2}), we derive $\hat{\cal H}$ up to order $a$ below.
First, $\hat X_I$ in Eq. (\ref{XA}) is linear in $\nabla_\mu$ and $\nabla_\mu^*$, so that
\begin{alignat*}1
e^{-i\frac{k}{a}\cdot x}\hat X_Ie^{i\frac{k}{a}\cdot x}
&= X_I-i\delta_\mu X_I aD_\mu +O(a^2),
\end{alignat*}
where $X_I=X_I(k)$ is calculated as $X_\mu=(-it/2)i(K_\mu+K_\mu^*)=t\sin k_\mu$ ($\mu=1,\cdots,d$) and 
$X_{d+1}=m+(b/2)\sum_\mu i(K_\mu-K_\mu^*)=m+b\sum_\mu(\cos k_\mu-1)$.
Then, we readily have
\begin{alignat*}1
e^{-i\frac{k}{a}\cdot x}\hat{\cal H}e^{i\frac{k}{a}\cdot x}
&=\gamma_I X_I-i\gamma_I\delta_\mu X_I aD_\mu  +O(a^2),\\
e^{-i\frac{k}{a}\cdot x}\hat{\cal H}^2e^{i\frac{k}{a}\cdot x}
&=X^2+\hat O 
-\gamma_{IJ}\delta_\mu X_I\delta_\nu X_J a^2F_{\mu\nu}+O(a^3),
\end{alignat*}
where we have defined $X^2=X_I^2$, 
$\gamma_{IJ}=[\gamma_I,\gamma_J]/4$, $\hat O=-2iX_I\delta_\mu X_I aD_\mu-\delta_\mu X_I\delta_\nu X_IaD_\mu aD_\nu$,
and 
\begin{alignat*}1
F_{\mu\nu}(x)=[D_\mu,D_\nu]=\partial_\mu A_\nu-\partial_\nu A_\mu,
\end{alignat*}
is the field strength of the external fields.

The smooth part $\bar\eta$ in Eq. (\ref{Eta2}) is the contribution of order $a^d$:
We compute the following in the limit $a\rightarrow0$:
\begin{widetext}
\begin{alignat*}1
\frac{\bar\eta}{a^d}
&=\frac{1}{2a^d}\sum_x\int_{-\pi}^\pi \frac{d^dk}{(2\pi)^{d}}
{\rm tr}\frac{\gamma_I X_I-i\gamma_I\delta_\mu X_I aD_\mu }{\sqrt{X^2+\hat O 
-\gamma_{IJ}\delta_\mu X_I\delta_\nu X_J a^2F_{\mu\nu}}}.
\end{alignat*}
Using the series expansion $\frac{1}{\sqrt{1-x}}=\sum_{n=0}^\infty\frac{(2n-1)!!}{n!2^n}x^n$ and
${\rm tr}~\gamma_{I_1}\gamma_{I_2}\cdots\gamma_{I_{d+1}}=(2i)^{d/2}\epsilon_{I_1I_2\cdots I_{d+1}}$, 
we have
\begin{alignat}1
\frac{\bar\eta}{a^d}&=
\frac{1}{2}\sum_x\int_{-\pi}^\pi \frac{d^dk}{(2\pi)^{d}}\frac{(2n-1)!!}{n!2^{n}X^{2n+1}}
{\rm tr}\gamma_I X_I \left(\gamma_{IJ}\delta_\mu X_I\delta_\nu X_J F_{\mu\nu}\right)^n +O(a)
\nonumber\\
&=(-1)^{n+1}c_n
\frac{i^n}{n!(2\pi)^n2^n}
\sum_x 
\epsilon_{\mu_1\nu_1\cdots\mu_n\nu_n}
F_{\mu_1\nu_1}\cdots F_{\mu_{n}\nu_{n}},
\label{SmoEta}
\end{alignat}
where we have set $d=2n$ to consider even $d$, and 
\begin{alignat}1
c_n\equiv\frac{(2n-1)!!(-1)^{n+1}}{2(2\pi)^n}
\int_{-\pi}^\pi d^dk
\frac{\epsilon_{II_1\cdots I_{d}}}{X^{2n+1}}
X_I\delta_{1}X_{I_1}\cdots\delta_{d}X_{I_{d}},
\label{CheNum}
\end{alignat}
\end{widetext}
is the $n$-th Chern number.
Equation (\ref{SmoEta}) has been derived in Ref. \cite{Qi:2008aa}.

\subsection{Chern number associated with Berry connection}
\label{s:CNBC}
We show that Eq. (\ref{CheNum}) is the Chern number associated with the many-body ground state.
The Chern number is defined as
\begin{alignat}1
c_n=\frac{1}{n!}\left(\frac{i}{2\pi}\right)^n\int{\rm tr}f^n,
\label{CheNumDef}
\end{alignat}
where $f$ is the Berry curvature 2-form  $f=\delta a+a^2$
defined using the Berry connection 1-form  $a=\psi^\dagger \delta \psi$.
Here, $\psi$ is the multiplet wavefunction of the negative energy states and 
$\delta$ is the external derivative with respect to $k_\mu$:
$\delta g= (\partial g/\partial k_\mu)d k_\mu=\delta_\mu gd k_\mu$.
It is known that 
\begin{alignat*}1
{\rm tr}f^n={\rm tr}\left(P(\delta P)^{2}\right)^n ={\rm tr}P(\delta P)^{2n} ,
\end{alignat*}
where $P$ is the projection operator to the ground state, $P=(1-{\cal H}/X)/2$ 
with ${\cal H}(k)=\gamma_IX_I(k)$.
Then,
\begin{alignat*}1
{\rm tr}f^n&=\frac{-1}{(2X)^{2n+1}}{\rm tr}\gamma_IX_I\gamma_{I_1}\delta X_{I_1}\cdots \gamma_{I_d}\delta X_{I_d}
\nonumber\\
&=\frac{-(2i)^n(2n)!d^dk}{(2X)^{2n+1}}\epsilon_{II_1\cdots I_d}X_I\delta_1 X_{I_1}\cdots\delta_d X_{I_d}.
\end{alignat*}
Substituting this into Eq. (\ref{CheNumDef}), we arrive at Eq. (\ref{CheNum}).

\subsection{A strategy for strong fields}
\label{s:SSF}
Thus far, we have derived $\eta$ in the $a\rightarrow0$ limit, which 
also implies the weak field limit.
Indeed, $c_n$ in Eq. (\ref{CheNumDef}) is given by $\psi$ or $P$ in the 
zero field limit. However, as shown by L\"uscher,
\cite{Luescher:1999fk}
$\eta$ on the lattice is generically given by the form of Eq. (\ref{SmoEta}),  and 
thus, we expect that 
the effects of finite $a$ and strong fields simply renormalize $c_n$ and $\eta_{\rm d}$.
Here, the topological nature of $\eta$ should constrain $c_n$ to be an integer.
In what follows, we regard $c_n$ as an unknown integer and determine it by 
the numerical computation of $\eta$.

\section{Two-dimensional HWD system}
\label{s:TDHWDS}
\begin{figure}[htb]
\begin{center}
\includegraphics[scale=1]{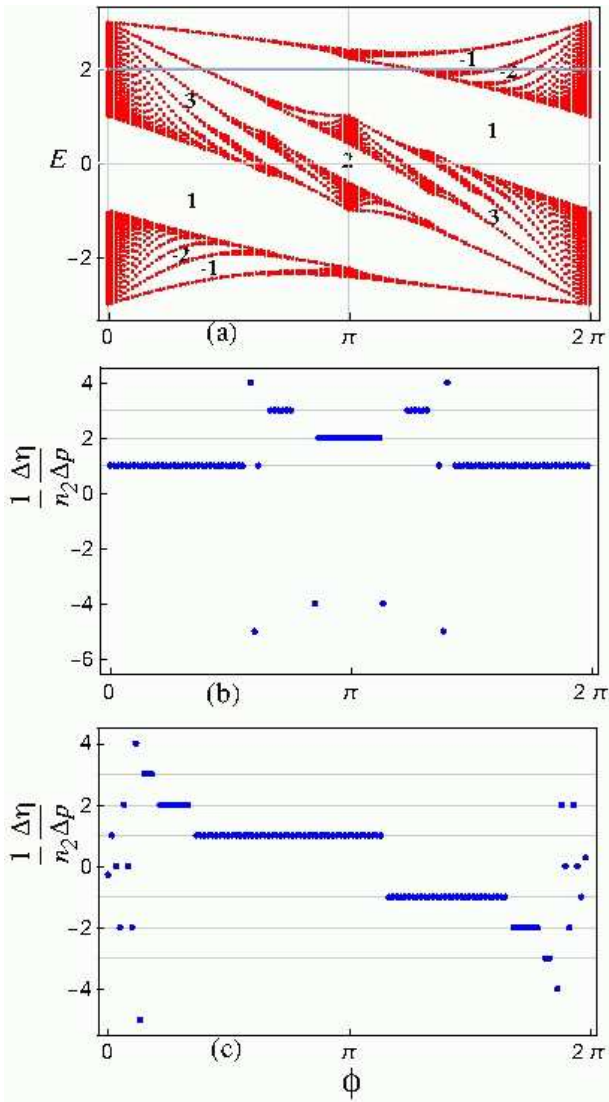}
\caption{
HWD model in $d=2$. The parameters used are $t=1$, $m=1$, $b=1$, and $q=120$.
The system size in $x_2$ is $n_2=120$.
(a) is the spectrum as a function of the magnetic flux per plaquette, $\phi$. Several Chern numbers 
for the occupied states below the gap are indicated.
(b) and (c) are computed using the Streda formula $(\Delta\eta/\Delta p)/n_2$ in Eq. (\ref{StrFor2D2}) 
at zero energy and at the energy $\mu=2$, respectively.
}
\label{f:2D}
\end{center}
\end{figure}

Let us consider a finite system under the periodic boundary condition 
in a uniform magnetic field $B$.
We assume a commensurate magnetic flux per plaquette, 
\begin{alignat}1
Ba^2\equiv\phi=\frac{2\pi p}{q},\quad p=0,\cdots,q ,
\label{MagFlu}
\end{alignat}
and consider the spectral flow as a function of 
$p$ with a fixed $q$.
This is the famous Hofstadter butterfly. 
\begin{figure}[htb]
\begin{center}
\begin{tabular}{cc}
\includegraphics[width=0.46\linewidth]{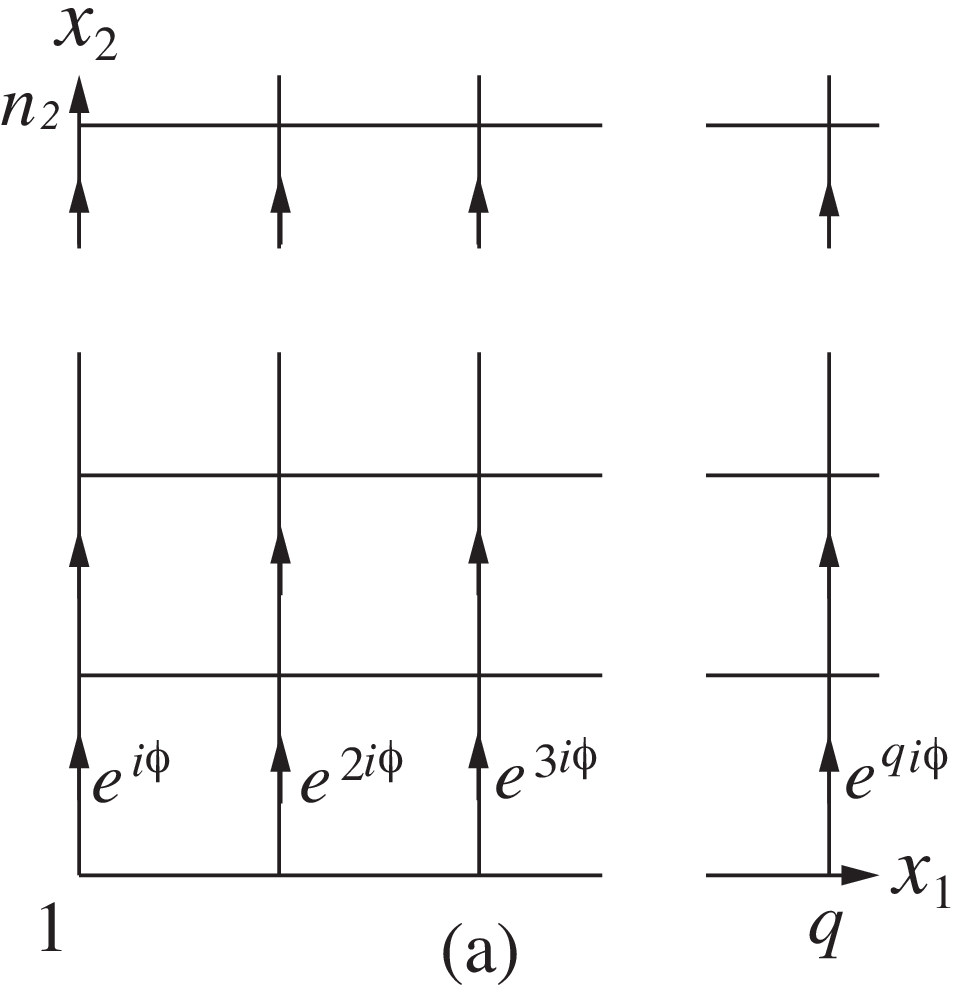}&
\includegraphics[width=0.46\linewidth]{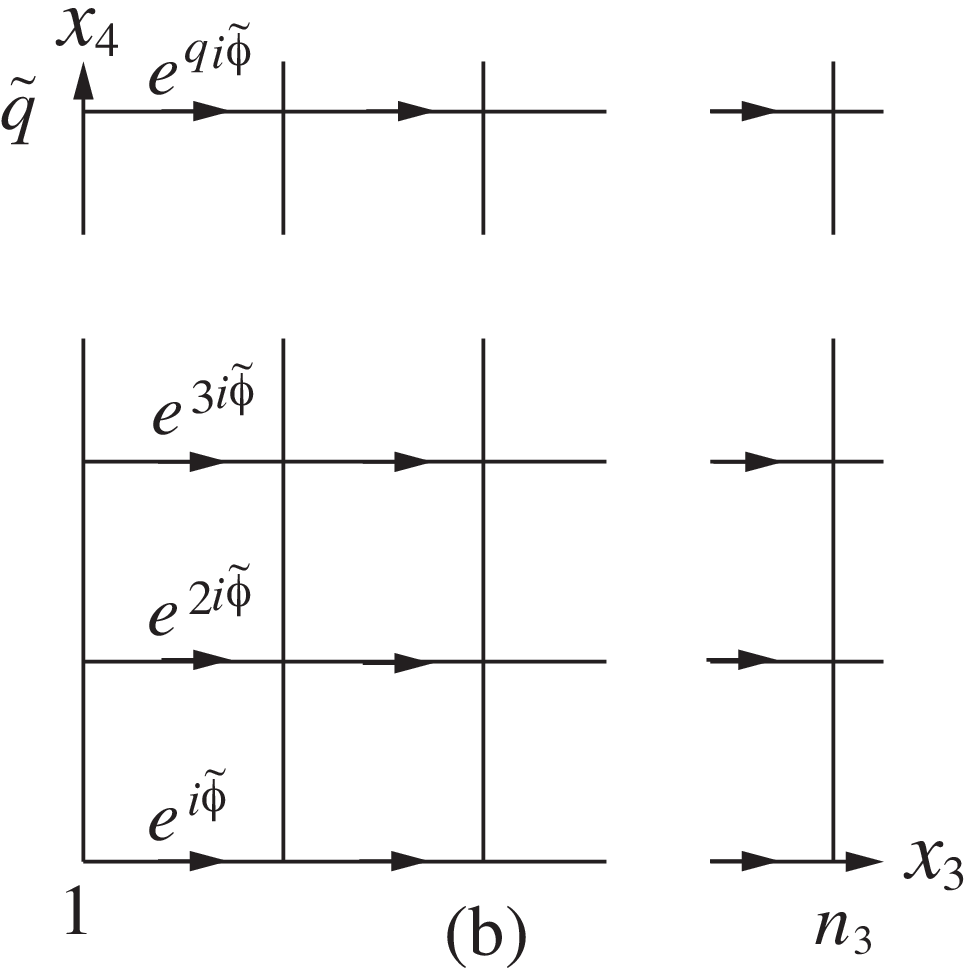}
\end{tabular}
\caption{
Landau gauge for a magnetic field (a) and an electric field (b). 
The system sizes toward $x_1$ and $x_4$ are $q$ and $\tilde q$, whereas they are
$n_2$ and $n_3$ toward $x_2$ and $x_3$, respectively.
The periodic boundary condition is imposed.
}
\label{f:lat}
\end{center}
\end{figure}
We take the Landau gauge
as depicted in Fig. \ref{f:lat}(a). 
The Hofstadter butterfly is shown in Fig. \ref{f:2D}(a) for the model whose
ground state has $c_1=1$ when $\phi=0$.
Several other Chern numbers
computed directly using the Berry curvature \cite{Thouless:1982uq,kohmoto:85,FHS05} are also shown.

$\bar\eta$ is given by (\ref{SmoEta}) with $iF_{12}=B$: We find 
\begin{alignat*}1
\bar\eta
&=\frac{c_1}{2\pi}\sum_xBa^2=\frac{c_1}{2\pi}Ba^2\cdot qn_2=c_1pn_2,
\end{alignat*}
where $qn_2$ is the number of the plaquettes on the plane.
Substituting this into Eq. (\ref{EtaTwo}), we reach
\begin{alignat}1
\eta=\frac{c_1}{2\pi}Ba^2\cdot qn_2+\eta_{\rm d}
=c_1pn_2+\eta_{\rm d}.
\label{Asy2D}
\end{alignat}
To see that this leads to the Streda formula, we note 
that the density of states below zero energy is $n_- \equiv N_-/(qn_2 a^2)$.
We also note that $\eta_{\rm d}$ is a function of $B$, but it should be constant if the energy gap is open 
at zero energy: When the gap closes and spectral flow occurs across zero energy, 
it discontinuously changes by an integer. 
From Eq. (\ref{Eta}), we have $\eta=N/2-N_-$, where $N$ is the total number of states. 
Thus, we reach
\begin{alignat}1
\left.\frac{d n_-}{d B}\right|_{\rm smooth}=-\frac{c_1}{2\pi}.
\label{StrFor2D}
\end{alignat}
This corresponds to the Streda formula for the Wilson-Dirac model.
For the practical numerical computations, it is convenient to 
regard $\eta$ as a function of $p$ with $q$ fixed. Then, Eq. (\ref{Asy2D}) is converted into 
\begin{alignat}1
\frac{1}{n_2}\frac{\Delta \eta}{\Delta p}\equiv\frac{\eta(p+1)-\eta(p)}{n_2} 
=c_1+\Delta\eta_{\rm d}' ,
\label{StrFor2D2}
\end{alignat}
where $\Delta\eta_{\rm d}'\equiv \Delta\eta_{\rm d}/n_2$ is obviously
zero when the Fermi energy is in the bulk gap.

In Fig. \ref{f:2D}(b), we show the l.h.s of Eq. (\ref{StrFor2D2}) 
computed directly from the numerical results of $N_\pm$ in (a).
We see several flat regions when the zero energy is in the bulk gap. From the r.h.s, 
it turns out that these are just the first Chern numbers of the ground state. 
Indeed, they are consistent with the Chern numbers indicated in Fig. \ref{f:2D}(a). 
On the other hand, if the gap closes, 
Eq. (\ref{StrFor2D2}) rapidly jumps. This is due to $\Delta\eta_{\rm d}'$. 

\begin{figure}[htb]
\begin{center}
\includegraphics[scale=1]{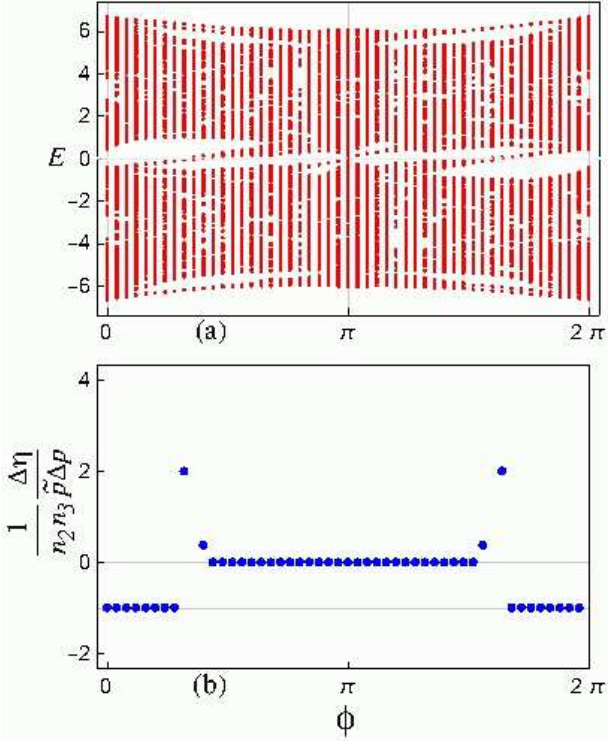}
\caption{
HWD model in $d=4$. The parameters used are
$t=1$, $m=1$, $b=1$, $q=50$, and $\tilde q=6$.
The numbers of sites along $x_2$ and $x_3$ are $n_2=n_3=4$.
(a) is the spectrum as a function of the magnetic flux $\phi$ for $\tilde\phi=\pi/3$ ($\tilde p=1$) fixed. 
(b) is the Streda formula in Eq. (\ref{Str4DNum}). 
}
\label{f:4D1}
\end{center}
\end{figure}

\begin{figure}[htb]
\begin{center}
\includegraphics[scale=1]{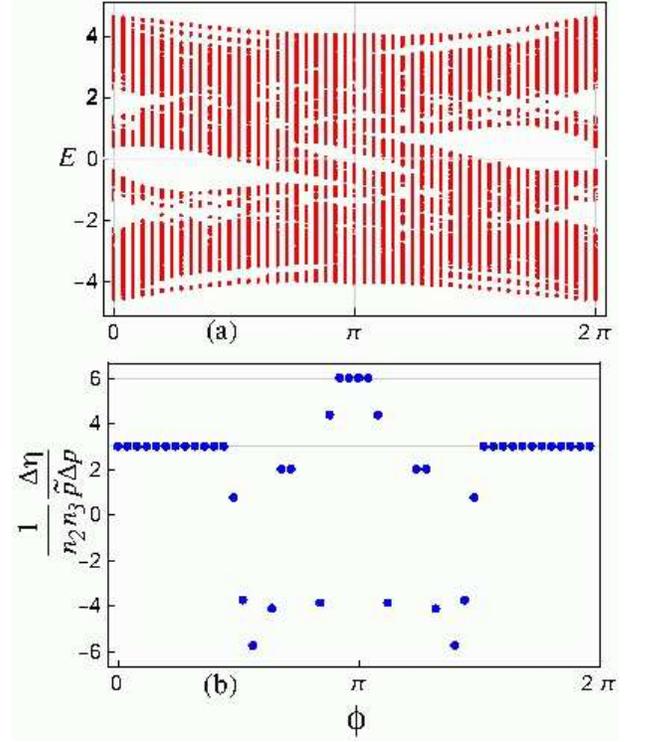}
\caption{
HWD model in $d=4$. The parameters used are the same as in Fig. \ref{f:4D1} except
$m=3.1$.
(a) is the spectrum. 
(b) is the Streda formula.
}
\label{f:4D2}
\end{center}
\end{figure}

For reference, we compute the model including the chemical potential $-\mu \sum_jc_j^\dagger c_j$ 
in the Hamiltonian.
In this case, $\eta$ can be computed by redefining $N_\pm$ as 
the numbers of states above and below the energy $\mu$. 
We show the result in Fig. \ref{f:2D}(c). It is consistent with the Chern numbers indicated in (a).
The Streda formula Eq. (\ref{StrFor2D}) or (\ref{StrFor2D2}) 
seems valid for not only zero energy but also finite energies, as it should be, 
since 
the original Streda formula \cite{Streda:1982aa} is quite generic.


\section{Four dimensional HWD system}
\label{s:FDHWDS}
In addition to a uniform magnetic field $B$ in Fig. \ref{f:lat}(a),
we introduce a uniform electric field $E$ in the $x_3$ direction, as depicted in Fig. \ref{f:lat}(b).
Note that $x_4$ is the imaginary time, and thus, the present Wilson-Dirac model is 
defined in the Euclidean space  in order for the Hamiltonian to be hermitian. 
We assume a commensurate electric field per plaquette, 
\begin{alignat}1
Ea^2\equiv\tilde\phi=\frac{2\pi \tilde p}{\tilde q},\quad \tilde p=0,1,\cdots,
\label{EleFlu}
\end{alignat}
as well as the magnetic field 
$\phi$ as in Eq. (\ref{MagFlu}), and consider the spectral flow as a function of magnetic field.

First, let us derive the Streda formula for the four-dimensional HWD model.
Setting $iF_{12}=B$ and $iF_{34}=E$ in Eq. (\ref{SmoEta}), we have
\begin{alignat*}1
\bar\eta
&
=-\frac{c_2}{(2\pi)^2}\sum_xBEa^4
=-\frac{c_2}{(2\pi)^2}BEa^4\cdot qn_2n_3\tilde q .
\end{alignat*}
This is nothing but the chiral anomaly term $\propto \bf B\cdot\bf E$.
Together with Eqs. (\ref{MagFlu}) and (\ref{EleFlu}), we reach
\begin{alignat*}1
\eta
&=-c_2p\tilde p n_2n_3+\eta_{\rm d}.
\end{alignat*}
The density of state below zero energy is given by $n_-\equiv N_-/(a^4qn_2n_3\tilde q)$, so that we have
\begin{alignat}1
\left.\frac{\partial n_-}{\partial (BE)}\right|_{\rm smooth}=\frac{c_2}{(2\pi)^2}.
\label{Str4D}
\end{alignat}
This can be regarded as the Streda formula in four dimensions describing the relationship between 
the density of states and 
the topological invariant $c_2$.
For the numerical calculations, 
as a function of $p$ and $\tilde p$ with $q$ and $\tilde q$ fixed, $\eta(p\tilde p)$, we have
\begin{alignat}1
\frac{1}{n_2n_3}\frac{\Delta\eta}{\Delta (p\tilde p)}=
\frac{1}{n_2n_3}\frac{1}{\tilde p}\frac{\Delta\eta}{\Delta p}
=-c_2+\Delta\eta_{\rm d}' ,
\label{Str4DNum}
\end{alignat}
where the first equality means that we compute the difference of $\eta$
with respect to $p$ with $\tilde p$ fixed.
Other notations are similar to those in Eq. (\ref{StrFor2D2}).

In Figs. \ref{f:4D1}(a) and \ref{f:4D2}(a),  
we show the spectra of the HWD model as functions of the magnetic flux $\phi$
with a small electric flux $\tilde\phi$ fixed.
Near $\phi\sim0$, the gaps at zero energy in \ref{f:4D1}(a) and \ref{f:4D2}(a) open, 
which are known to have $c_2=1$ and $-3$,
respectively, at $\phi=0$. \cite{1126-6708-2002-09-025,Qi:2008aa}
We expect that even with a magnetic field, the gaps keep the
same Chern numbers until the gaps close in a strong magnetic field regime.
To see this, we show the numerical calculations of the l.h.s of Eq. (\ref{Str4DNum})
in Figs. \ref{f:4D1}(b) and \ref{f:4D2}(b). 
Indeed, there appear flat regions with the same Chern numbers.
Therefore, we expect that the Streda formula (\ref{Str4D}) or (\ref{Str4DNum}) is valid 
even in a strong magnetic field. Interestingly, 
we find another gapped ground state with $c_2=-6$ around a very strong magnetic field $\phi=\pi$
in Fig. \ref{f:4D2}(b).

\begin{figure}[htb]
\begin{center}
\includegraphics[scale=.8]{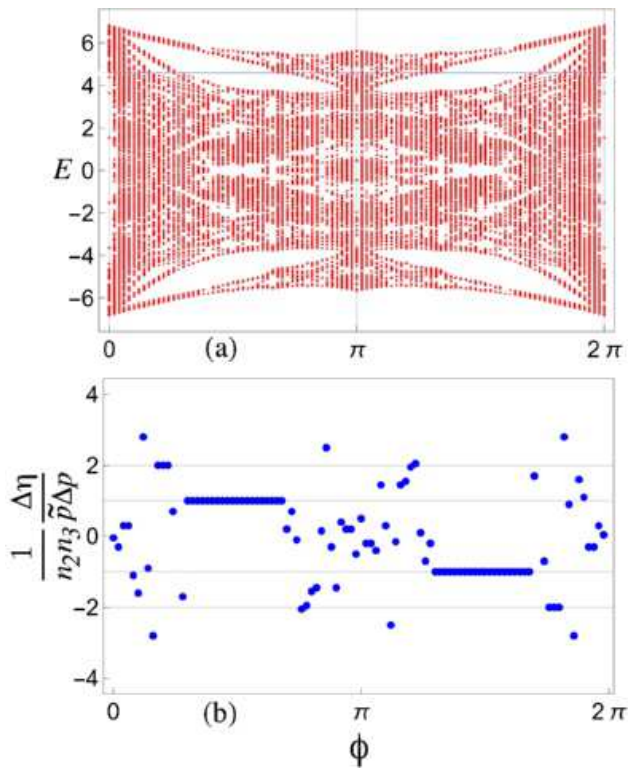}
\caption{
A tight-binding  model in $d=4$. 
The parameters used are $t_\mu=1$ (uniform hopping), $q=100$, and $\tilde q=16$.
The system size is $n_2=n_3=10$.
(a) is the spectrum for $\tilde\phi=\pi/2$ ($\tilde p=4$) fixed. 
(b) is the Streda formula in Eq. (\ref{Str4DNum}) with chemical potential $\mu=4.6$.
For reference, we show a horizontal thin line at energy $\mu=4.6$ in (a). 
}
\label{f:4DTB}
\end{center}
\end{figure}

\section{Generic systems: A conjecture}
\label{s:GSC}
What we have learned in the QHE is that a simple tight-binding model, which is 
topologically trivial,
becomes nontrivial once a magnetic field is switched on and
a single band spectrum separates into many Landau levels with finite Chern numbers.
This motivates us to investigate a simple tight-binding model in four dimensions,
\begin{alignat}1
H=-\sum_{\mu=1}^4t_\mu\sum_j\left(e^{-i\phi_{\mu,j}}c_j^\dagger c_{j+\hat\mu}+{\rm h.c.}\right) ,
\label{HamTB}
\end{alignat}
where the magnetic and electric fields are introduced as in Fig. \ref{f:lat}. 
Here, remember that
the Streda formula in Eq. (\ref{StrFor2D}) or (\ref{StrFor2D2}) is derived for the HWD model in $d=2$, but
it holds more generically in a model-independent manner, as shown by Streda.\cite{Streda:1982aa}
Thus, we assume that the Streda formula in Eq. (\ref{Str4D}) or (\ref{Str4DNum}) 
is also valid for more generic models in $d=4$,
and apply it to the simple tight-binding model in Eq. (\ref{HamTB}).

We show in Fig. \ref{f:4DTB}(a) the spectrum of the model (\ref{HamTB})
as a function of the magnetic flux. 
Although the spectrum is gapless at zero energy, we find several gap structures at $E\sim\pm 5$.
Let us compute the Streda formula Eq. (\ref{Str4DNum}) at energy $\mu=4.6$, which is 
shown in Fig. \ref{f:4DTB}(b).
At four gaps from $\phi=0$ to $2\pi$,  $c_2$ reads
$-2,-1,1,2$.

One may wonder if there is room for the first Chern number
to characterize the numerically observed flat regions in Fig. \ref{f:4DTB},
since
at least in the case $t_3, t_4\ll t_1,t_2$, the model can be only a layered square lattice system. 
Let us consider the extreme case, $t_3=t_4=0$.
In this case, the Landau levels should be characterized by $c_1$, since the system is copies of 
independent two-dimensional systems, and Landau levels obviously carry the first Chern numbers, 
not the second Chern numbers.
Note that the number of copies is $n_3\tilde q$. Thus,  
Eq. (\ref{Asy2D}) is modified as $\eta=c_1pn_2\cdot n_3\tilde q+\eta_{\rm d}$.
Therefore, we expect in this case,
\begin{alignat}1
\frac{1}{n_2n_3}\frac{1}{\tilde q}\frac{\Delta\eta}{\Delta p}
=c_1+\Delta\eta_{\rm d}' .
\label{Str4D2D}
\end{alignat}
The differences in $\tilde p$ and $\tilde q$ in the denominators of the l.h.s in Eqs. (\ref{Str4DNum}) and
(\ref{Str4D2D})
should be noted.
If we interpret the observed values of the flat regions in Fig. \ref{f:4DTB} 
to be $c_1$ as the result of the two-dimensionality, 
Eq. (\ref{Str4D2D}) tells that $c_1=-c_2\cdot\tilde p/\tilde q=-c_2/4=1/2,1/4,-1/4,-1/2$ from $\phi=0$ to $2\pi$.
Thus, these 
cannot be the first Chern numbers. 
If we reduce the values of $t_3$ and $t_4$, say, up to the order of $t_1/10,t_2/10$, the spectrum 
almost reproduces the two-dimensional butterfly of the square lattice system, and the first Chern numbers 
computed using
Eq. (\ref{Str4D2D}) are consistent with those of the square lattice system.
Therefore, it is quite natural to conclude that the uniform tight-binding model in four dimensions
has a nontrivial gapped ground state characterized by $c_2$. 

\section{Summary and discussion}
\label{s:SD}

We have explored the chiral anomaly of the Wilson-Dirac model on the lattice in strong external fields.
Taking the weak field limit for the spectral asymmetry, 
we have rederived the chiral anomaly, paying special attention to the 
Chern number due to the Berry connection.
In two dimensions, the relationship between the spectral asymmetry and the chiral anomaly 
is the same as the Streda formula.
Thus, the generalized Streda formula we have derived in this paper enables us to compute the Chern numbers
using eigenvalues only, without using eigenvectors.
The results of several numerical calculations have suggested that the generalized Streda formula is 
valid for not only the Wilson-Dirac model but also the simple tight-binding model.

It is natural that the Landau level of a system in $2n$ dimensions allows the $n$th Chern numbers.
However, there is room for lower Chern numbers, reflecting the fact that a $2n$-dimensional system can be
a layered system in lower dimensions, at least in some limit,
as we have mentioned in Sect. \ref{s:SD}. The dimensionality and the order of the Chern number 
may be an interesting future issue.
The direct derivation of the Streda formula in higher dimensions, especially in four dimensions for more generic
systems, should also be addressed, since the second Chern number is relevant to a recent interesting topic,
i.e., the observation of the chiral anomaly in a crystal.

\acknowledgements
This work was supported in part by Grants-in-Aid for Scientific Research Numbers 
25400388, 26247064, and 24540247
from the Japan Society for the Promotion of Science.

\bibliography{67888}

\end{document}